\documentclass[12pt]{article}
\usepackage{graphicx,psfrag,epsf}
\usepackage{natbib}
\usepackage{color}
\usepackage{bbm}     %for the indicator function (but font type 3)
\usepackage{dsfont}  %for the indicator function (font type 1)
\usepackage[centertags]{amsmath}
\usepackage{booktabs}
\usepackage{adjustbox}
\usepackage{multirow}
\usepackage{caption}
\usepackage{rotating}
\usepackage{hyperref}
\usepackage{amssymb}
\def\R{\mathds{R}}
\def\N{\mathds{N}}
\def\bs{\boldsymbol}

\newcommand{\indicator}[1]{\mathds{1}_{\left[ {#1} \right] }}

\def\z{{\bf z}}

\def\x{{\bf x}}
\def\X{{\bf X}}
\def\y{{\bf y}}
\def\Y{{\bf Y}}

\def\argmin{\mathop{\rm arg\,min}\limits}

\newtheorem{theorem}{Theorem} %[section]

\title{The quantile-based classifier with variable-wise parameters}
\author{Marco Berrettini$^1$\thanks{corresponding author: marco.berrettini2@unibo.it}, Christian Hennig$^1$ and Cinzia Viroli$^1$\\
$^1$ Deparment of Statistical Sciences, University of Bologna}

\date{}

\begin{document}
\maketitle

\begin{abstract}
Quantile-based classifiers can classify high-dimensional observations by minimising a discrepancy of an observation to a class based on suitable quantiles of the within-class distributions, corresponding to a unique percentage for all variables. The present work extends these classifiers by introducing a way to determine potentially different optimal percentages for different variables. Furthermore, a variable-wise scale parameter is introduced. A simple greedy algorithm to estimate the parameters is proposed. Their consistency in a nonparametric setting is proved. Experiments using artificially generated and real data confirm the potential of the quantile-based classifier with variable-wise parameters.
\end{abstract}
\noindent%

\section{Introduction}

Quantiles can be useful for both supervised and unsupervised classification.
They can be used to focus on tails of the class conditional distributions rather than only their central moments.
For supervised classification, \cite{HV2016} proposed a quantile-based classifier, referred to as (original) quantile classifier (OQC), that considers a sum of variable-wise discrepancies of the components of a new observation (to be classified) to the within-class univariate quantiles. The method is a generalization of the centroid-based \citep{THNC02,THNC03} and the median-based \citep{HTX09} classifiers.
It has been shown that the strategy works quite well for skewed data and heavy tails compared to other classifiers. A recent extension based on a regularised loss function \citep{LM20} performs comparatively well in different scenarios, especially for limited sample sizes or dimensionality.

\cite{HV2016} included a somewhat surprising empirical result, namely that using the same percentage for defining quantiles for all variables (only involving a check whether a variable is left- or right-skewed) worked better than any approach explored at the time to choose different percentages for different variables, even where these variables had different distributional shapes.

\cite{quclust} developed a method for unsupervised classification based on quantiles, involving an idea for determining variable-wise percentages that can be adapted to unsupervised classification. In this paper we use a similar approach to propose a generalised classifier with variable-wise percentages and different weights.
% The resolution of the mystery of the optimality of a single percentile is inspired by the work on quantile-based clustering \citep{quclust}, where an unsupervised classification strategy based on quantiles is introduced along the lines of classical $K$-means.
Here we show that the proposed quantile classifier with variable-wise percentage parameters can outperform the earlier version with a common quantile for all variables.
Numerical studies provide very good results compared to a number of popular classification methods.

In Section \ref{squan}, we introduce the original quantile classifier and describe its extension by introducing variable-wise percentage and scale parameters. We also address the computation of the new classifier. In Section \ref{scons} we prove in a general nonparametric situation that the parameter estimates employed by the new classifier are consistent for their corresponding population versions. Section \ref{sexp} compares the proposed quantile classifier to several other popular classifiers based on simulated and real data sets. Section \ref{sconc} concludes the paper.

\section{The quantile classifier}\label{squan}
In this section we describe the quantile classifier and its main properties.

Let $\Pi_1$ and $\Pi_2$ be two populations with probability densities $F_1$ and $F_2$ on $\mathbb{R}^p$.
Quantile-based classifiers \citep{HV2016} assign a new observation $\textbf{z}=(z_1,\ldots,z_p)\in\mathbb{R}^p$ to the population from which it has lowest quantile discrepancy, as defined below, along all variables. Formally, $\textbf{z}$ is allocated to $\Pi_1$ if
\begin{eqnarray}\label{e:decision}
\sum_{j=1}^p \{\Phi_{2jn}(\z,\theta)-\Phi_{1jn}(\z,\theta)\}>0,
\end{eqnarray}
and otherwise to $\Pi_2$. Here, the quantile discrepancy $\Phi_{kjn}$ is
defined for any variable $j=1,\ldots,p,$ classes $k=1, 2$, and a percentage $\theta$, as
\begin{eqnarray}\label{e:discrepancy}
\Phi_{kjn}(\z,\theta)
&=&\left(\theta+(1-2\theta)\indicator{z_j\le q_{kjn}(\theta)}\right)|z_j-q_{kjn}(\theta)|,
\end{eqnarray}
where $q_{kjn}(\theta)$ is the empirical marginal quantile function of variable $j$ within class $k$ evaluated at $\theta \in (0,1)$, i.e., the sample $\theta$-quantile. The index $n$ indicates an underlying sample of $n_1+n_2=n$ training observations, $n_1$ of which stem from $\Pi_1$, and $n_2$ stem from $\Pi_2$.

The quantile function for a one-dimensional distribution $P$ with cdf $F$ is defined as
$q_P(\theta)=\inf\{x:F(x)\geq \theta\}$. This is the infimum of the not necessarily unique values of $\xi$ that minimises the following variability measure,
\begin{eqnarray}\label{eqn:qdist.centroid.0}
\theta\int_{x>\xi}|x-\xi|dP(x)+(1-\theta)\int_{x< \xi}|x-\xi|dP(x).
\end{eqnarray}
Since for $\theta=0.5$ the quantile discrepancy is twice the $L_1$-distance from the median, the decision rule (\ref{e:decision}) includes the median classifier \citep{HTX09} when $\theta=0.5$.
For all values $\theta$, the quantile discrepancy not being based on squares shares with the $L_1$-distance its better
resistance against outliers compared to Euclidean $L_2$-distance.

\citep{HV2016} propose to choose the $\theta$-value as optimal that produces the optimal correct classification rate within the training data.
\subsection{Variable-wise $\theta$ and a scale parameter}
The original quantile classifier as defined above is based on finding
a single optimal $\theta$ for all variables simultaneously. Considering that each variable might have different skewness, tail rate and shape, it appears to be promising to choose variable-wise values $\theta_1,\ldots,\theta_p$. Using them, the new observation $\textbf{z}$ can be allocated to population $\Pi_1$ if
\begin{eqnarray}\label{e:decision2}
\sum_{j=1}^p \{\Phi_{2jn}(\z,\theta_j)-\Phi_{1jn}(\z,\theta_j)\}>0.
\end{eqnarray}
\cite{HV2016} report on having tried out several strategies for choosing variable-wise $\theta$, but  all these resulted in a worse classification performance on independent data than just choosing a single $\theta$. The problem with variable-wise $\theta_j$ is that plugging different $\theta_j$ for different variables into (\ref{e:discrepancy})) will induce relative weights $\left(\theta_j+(1-2\theta_j)\indicator{z_j\le q_{kjn}(\theta_j)}\right)$ for the different variables, regardless of their information content for classification. In particular, variables with $\theta_j$ close to 0 or 1 will have more influence than other variables regardless of their effective discriminative power.
Instead, \cite{HV2016} proposed to set the within-class skewness averaged over classes of all variables in the same direction by just multiplying variables that have the wrong skewness direction by -1, so that the information of the different variables for choosing the joint $\theta$ becomes more compatible. This preprocessing step resulted in good classification performances.

But this is a very rough adjustment, and in the present work we show that there is a sensible way to choose variable-wise $\theta$s together with re-scaling the variables in such a way that the contributions of different variables are still properly balanced.

From now on we consider $K$ classes where not necessarily $K=2$. The decision rule in (\ref{e:decision}) can be easily generalised to $K>2$ classes by just allocating an observation $\z$ to the population which gives the lowest quantile discrepancy $\sum_{j=1}^p \Phi_{kjn}(\z,\theta)$, with $k=1,\ldots,K$.

Given an observed sample $\tilde \y_n=(\y_1,\ldots,\y_n),$ where $\y_i=(\x_i,c_i),\ i=1,\ldots,n, c_i\in\{1,\ldots,K \}$ being the known number of the class to which $\x_i\in \R^p$ belongs, the aim is to find first the optimal $\theta_j,\ j=1,\ldots,p$, which minimise the sum of variable-wise quantile discrepancies
\begin{eqnarray}\label{e:rule}
\hat\theta= \argmin_{\theta_1,\ldots,\theta_p} {\sum_{i=1}^n \sum_{j=1}^p \Phi_{c_ijn}(\x_i,\theta_j)},
\end{eqnarray}
and then to allocate a new observation $\z$ according to the minimum $\sum_{j=1}^p\Phi_{kjn}(\z,\hat\theta_j)$ for all $k=1,\ldots,K$.
%Here $q_{kjn}(\hat\theta)$ in the definition of $\Phi_{kj}(\z,\hat\theta_j)$ refers to the sample quantile of the training data from class $k$.

In unsupervised quantile-based clustering, \cite{quclust} established the connection between the quantile discrepancy and the asymmetric Laplace distribution. More specifically, for $p=1$ and $K=1$, minimising the quantile discrepancy on a sample of data by choice of $\theta$ is equivalent to maximising the likelihood of the asymmetric Laplace distribution with $\lambda=1$, defined by
\begin{eqnarray}\label{density}
f(x;\lambda,\theta)=\lambda\theta(1-\theta)e^{- \lambda\left\{\theta+(1-2\theta)\indicator{x<q(\theta)} \right \}|x-q(\theta)|}.
\end{eqnarray}
$\lambda$ is a scale parameter, and its impact on the quantile classifier is that its introduction assigns a weight to the contribution of a variable. The general quantile-based classifier can be defined as the problem of assigning a new observation to the class minimising the discrepancy
$$\sum_{j=1}^p \lambda_j \Phi_{kjn}(\z,\theta_j),$$
where $\lambda_j $ and $\theta_j$ are estimated on the training set
$\tilde \y_n$ as
\begin{eqnarray}\label{e:rule2}
T_n(\tilde\y_n)=
\argmin_{\boldsymbol\lambda,\boldsymbol\theta}
\left(\sum_{i=1}^n \sum_{j=1}^p \lambda_j \Phi_{c_ijn}(\x_i,\theta_j) - n \sum_{j=1}^p \log \lambda_j \theta_j (1 - \theta_j)\right),
\end{eqnarray}
where $- n \sum_{j=1}^p \log \lambda_j \theta_j (1 - \theta_j)$ derives from the normalization constant of the density (\ref{density}). It can be interpreted as a penalty term that penalises $\theta_j$-values that are too close to 0 or 1, and small scale parameters $\lambda_j$. Let
\[
\Psi ( \boldsymbol\lambda,\boldsymbol\theta,\tilde\y_n)=\sum_{i=1}^n \sum_{j=1}^p \lambda_j \Phi_{c_ijn}(\x_i,\theta_j) - n \sum_{j=1}^p \log \lambda_j \theta_j (1 - \theta_j).
\]

\subsection{Estimation}
Parameter estimation can be obtained through a greedy algorithm that starts with an initialization step and then alternates between updating $\theta_j$ and $\lambda_j$ until convergence of the objective function $\Psi ( \boldsymbol\lambda,\boldsymbol\theta,\tilde\y_n)$ in (\ref{e:rule2}). To this aim we need the scores of the objective function with respect to the parameters of interest.

By equating the score of $\Psi ( \boldsymbol\lambda,\boldsymbol\theta,\tilde \y_n)$ to zero with respect to each scale parameter $\lambda_j$, we get an estimate in closed form given by
\begin{eqnarray}\label{est1}
\lambda_j=\frac{n}{\sum_{i=1}^n \Phi_{c_ijn}(\x_i,\theta_j)}.
\end{eqnarray}
The estimates of $\theta_j$ can be obtained similarly and are given in closed form by taking a root of the quadratic equation:
\begin{eqnarray}\label{est2}
\theta_j^2 \lambda_j \sum_{i=1}^n(x_{ij} -q_{c_ijn}(\theta_j)) -\theta_j \left(2n+ \lambda_j \sum_{i=1}^n(x_{ij} -q_{c_ijn}(\theta_j))\right)+n=0.
\end{eqnarray}

Thus, the algorithm is very fast. It is represented in the following scheme:

\medskip
\noindent\rule[0.5ex]{\linewidth}{1pt}
\begin{enumerate}
  \item \emph{Initialization}: For each variable, choose randomly a value $\theta_j$ and compute the class-conditional quantiles $q_{kjn}\left(\theta_{j}\right)$; set $\lambda_j=1$.
  \item Repeat the following until $\Psi ( \boldsymbol\lambda,\boldsymbol\theta)$  stops changing:
  \begin{enumerate}
    \item For $j=1,\ldots,p$ compute $\theta_j$ by equation (\ref{est1}).
    \item For $j=1,\ldots,p$ compute $\lambda_j$ by equation (\ref{est2}).
  \end{enumerate}
\end{enumerate}
\noindent\rule[0.5ex]{\linewidth}{1pt}

\medskip
We call the resulting classifier variable-wise quantile classifier (VWQC).
\section{Consistency}\label{scons}
We show here that in a nonparametric situation with data from classes $1,\ldots,K$ with probabilities $\pi_1,\ldots,\pi_K$ generated by distributions $Q_1,\ldots,Q_K$, respectively, for $n\to\infty$ the estimated parameters $(\bs\lambda,\bs\theta)$ converge almost surely (a.s.) to their population counterparts. This is somewhat different from the consistency result in \cite{HV2016}, where in the probabilistic limit a parameter was chosen that minimised the misclassification probability in the population. This was possible because the parameter $\theta$ was one-dimensional in \cite{HV2016}, and it was feasible to choose it minimising the in-sample misclassification probability. This is numerically too tedious here. Consequently, we can show that the class-wise distributions are ``asymptotically correctly'' represented in the classification rule, but not that this is asymptotically optimal for classification. Note that due to the nonparametric nature of the result, it is particularly not assumed that the variables are independent within classes, even though the classifier does not make use of potential dependence.

We consider the following setup. Let $(\X_i,C_i)_{i\in\N}$ be a sequence of i.i.d. random variables with ${\cal L}(\X_i,C_i)=Q$, where $Q\{C_i=k\}=\pi_k$ for $k\in 1,\ldots,K$, and
${\cal L}(\X_i|C_i=k)=Q_k$. Let $\tilde\Y_n=((\X_1,C_1),\ldots,(\X_n,C_n))$.
In the proof we will rely on uniform convergence and continuity. Sample quantiles are not continuous in $\theta$. Therefore for the theory we will consider the ``location'' of a within-class distribution ($q(\theta)$ in (\ref{density})) as additional parameter, even though this ultimately evaluates to the sample or population quantile by minimising (\ref{eqn:qdist.centroid.0}). Therefore define for $j=1,\ldots,p$:
\begin{eqnarray*}
\Phi_{j}(\z,\theta,\xi)
&=&\left(\theta+(1-2\theta)\indicator{z_j\le \xi}\right)|z_j-\xi|,\\
V_n(\bs\lambda,\bs\theta,\bs\xi,\tilde\Y_n)&=&\frac{1}{n}\sum_{i=1}^n \sum_{j=1}^p \lambda_j \Phi_{C_ijn}(\X_i,\theta_j,\xi_{C_ij}) - \sum_{j=1}^p \log \lambda_j \theta_j (1 - \theta_j),
\end{eqnarray*}
where $\bs\xi$ is the vector collecting all $\xi_{kj}$. $V_n$ is just $\Psi$ divided by $n$ for making laws of large numbers available. Therefore,
\begin{eqnarray*}
  T_n(\tilde\Y_n)&=&
\argmin_{\boldsymbol\lambda,\boldsymbol\theta,\bs\xi}V_n(\bs\lambda,\bs\theta,\bs\xi,\tilde\Y_n).
\end{eqnarray*}
Further let
\begin{eqnarray*}
V(\bs\lambda,\bs\theta,\bs\xi,Q)&=&\sum_{k=1}^K\pi_k\int \sum_{j=1}^p \lambda_j \Phi_{j}(\x,\theta_j,\xi_{cj})dQ(\x,c) - \sum_{j=1}^p \log \lambda_j \theta_j (1 - \theta_j),\\
T(Q) &=& \argmin_{\boldsymbol\lambda,\boldsymbol\theta,\bs\xi}V(\bs\lambda,\bs\theta,\bs\xi,Q).
\end{eqnarray*}
The consistency proof will rely on showing that parameter estimators for large $n$ do not leave a compact set, but (considering a single variable) $\lambda\to\infty$ and $\theta\to 0$ or $\theta\to 1$ may happen together without constraints on the parameter space, causing trouble with compactness. This can be avoided by either constraining $\theta_j\in[\tau,1-\tau],\ \tau>0$, or $\lambda_j\le \lambda^+<\infty$ for $j\in\{1,\ldots,p\}$. We will impose the latter constraint here. The parameter space used is
\[
S=\{(\bs\lambda,\bs\theta,\bs\xi):\ \theta_j\in (0,1), \lambda_j\in (0,\lambda^+],\ \xi_{kj}\in\R,\ j\in\{1,\ldots,p\}\},
\]
and the argmin in the definition of $T$ and $T_n$ is taken over $S$.

We require the following assumptions ($\|\bullet\|_{1}$ denotes the L1-norm):
\begin{description}
\item[A1] $B_k=\int \|\x\|_{1} dQ_k(\x)<\infty$ for $k=1,\ldots,K$.
\item[A2] $T(Q)$ is uniquely defined.
%\item[A3]
%\[
%\exists \epsilon, delta>0 \forall \{\bxi_1,\ldots,\bxi_K\}\subset \R^p:
%P\{\min_{k=1,\ldots,K}\|\x-\bxi_k\|>\epsilon\}>\delta.
%\]
\end{description}
\begin{theorem} \label{tcons}
If $(\X_1,C_1), (\X_2,C_2),\ldots \sim Q$ i.i.d., and assumptions A1 and A2 hold,
then, for $n\to\infty$:  $T_{n}(\tilde\Y_n)\to T(Q)$ a.s.
\end{theorem}
{\bf Proof:}
The principle of the proof follows the proof of Theorem 1 in
\cite{quclust} and is to show that
$T_{n}(\tilde\Y_n)$ for large enough $n$ has to lie in a compact set
${\cal C}$.
In this compact set, by the uniform law of large numbers,
$V_{n}(\bs\lambda,\bs\theta,\bs\xi,\tilde\Y_n)$ will converge uniformly to
$V(\bs\lambda,\bs\theta,\bs\xi,Q)$, which in turn, together with continuity,
will also enforce
the minimiser to converge. $(\bs\lambda,\bs\theta)$ optimizing $V_{n}$
are enforced to eventually lie in a compact set by the penalty term
$-\log \lambda\theta(1-\theta)$.

To show that $T_{n}(\tilde\Y_n)$ for large enough $n$ is a.s. in a compact set
${\cal C}$, first show that the minimum of $V_{n}$ is asymptotically bounded from above. This is then used to constrain $\bs\lambda, \bs\theta, \bs\xi$ into a compact set. Define $(\bs\lambda_0,\bs\theta_0,\bs\xi_0)$
as follows: For $j=1,\ldots,p,\ k=1,\ldots,K$:
$\theta_{0j}=\frac{1}{2},\ \lambda_{0j}=1,\ \xi_{0kj}=0$. Then,
\[
V_{n}(\bs\lambda_0,\bs\theta_0,\bs\xi_0,\tilde\Y_n)=
\frac{1}{n}\sum_{i=1}^n \sum_{j=1}^p \frac{1}{2}|X_{ij}| -p\log \frac{1}{4}.
\]
The first part converges a.s. to $B_1=\frac{1}{2}\sum_{k=1}^K \pi_kB_k<\infty$ by A1.

If $\lambda_j\to 0,\ \theta_j\to 1,$ or $\theta_j\to 0$, then $- \sum_{j=1}^p \log \lambda_j \theta_j (1 - \theta_j)\to\infty$, whereas $\frac{1}{n}\sum_{i=1}^n \sum_{j=1}^p \lambda_j \Phi_{C_ijn}(\X_i,\theta_j,\xi_{C_ij})$ is bounded from below by $-\frac{1}{n}\sum_{i=1}^n \sum_{j=1}^p|X_{ij}|$, which converges a.s. to $-B_1>-\infty$. This means that there is $\tau>0$ so that $T_{n}(\tilde\Y_n)=(\lambda_{1n},\ldots,\lambda_{pn},\theta_{1n},\ldots,\theta_{pn})$ must have $\lambda_{jn}\in[\tau,\lambda^+],\ \theta_{jn}\in[\tau,1-\tau]$ for $j=1,\ldots,p$.

For given $\bs\lambda,\bs\theta$, the class-wise sample quantiles belonging
to $\bs\theta$ are the values for $\bs\xi$ minimising $V_n(\bs\lambda,\bs\theta,\bs\xi,\tilde\Y_n)$; analogously the population quantiles minimise
$V(\bs\lambda,\bs\theta,\bs\xi,Q)$. \cite{Mas82} showed that for $\tau>0$ and $\theta\in[\tau,1-\tau]$, the sample quantiles converge uniformly a.s. to the
population quantiles. Therefore, for large enough $n$, $\bs\xi_n$ of $T_{n}(\tilde\Y_n)$ must be in a bounded neighbourhood of the set of quantile vectors of $Q_1,\ldots,Q_K$ for $\theta\in[\tau,1-\tau]$, and therefore $T_{n}(\tilde\Y_n)$ as a whole will be in a compact set a.s.

According to \cite{vdvaart98}, Example 19.8 (sometimes referred to as ``uniform law of large numbers''), if ${\cal F}=\{f_{\eta}:\ \eta\in {\cal C}\}$ is a set of measurable functions with $\eta\mapsto f_\eta(x)$ continuous for all $x$, ${\cal C}$ compact, and $\exists F\ge |f_\eta| \forall \eta\in {\cal C},\ \int F dP<\infty$, then
\[
\sup_{\eta\in{\cal C}} \left|\frac{1}{n}\sum_{i=1}^n f_\eta(\x_i)-\int f_\eta(\x)dP(\x)\right|\to 0\mbox{ a.s.}
\]
For fixed $x$,
$g(x,\theta, \xi)=\left\{\theta+(1-2\theta)\indicator{x<\xi} \right \}|x-\xi|$
is continuous in $(\xi,\theta)$, because
$\xi\to x\Rightarrow g(x,\theta, \xi)\to 0$ regardless of whether $\xi$
comes from above or from below.
Therefore, for fixed $\y=(x_1,\ldots,x_p,c)$,
\[
U(\bs\lambda,\bs\theta,\bs\xi,\y)=\sum_{j=1}^p \lambda_j g(x_{j},\theta_j,\xi_{cj})-\sum_{j=1}^p\log \left[\lambda_j(\theta_j(1-\theta_j))\right]
\]
is continuous. It can be bounded by a $Q$-integrable function. Let
$\infty>\xi^+>\max_{k\in\{1,\ldots,K\},\theta\in\{\kappa,1-\kappa\}}|q_k(\theta)|$,
$q_k(\theta)$ being the $\theta$-quantile of $Q_k$. Then,
\begin{eqnarray*}
U(\bs\lambda,\bs\theta,\bs\xi,\y)&\le& U^+(\y)=\sum_{j=1}^p \lambda^+(|x_j|+\xi^+) -\sum_{j=1}^p\log \left[\frac{\kappa^2}{2}\right],\\
\int U^+(\y) dQ(\y) &<&\infty \mbox{ because of A1.}
\end{eqnarray*}
Therefore,
\begin{equation}\label{eq:uloln}
\sup_{(\bs\lambda,\bs\theta,\bs\xi)\in{\cal C}}
|V_{n}(\bs\lambda,\bs\theta,\bs\xi,\tilde\Y_n)-V(\bs\lambda,\bs\theta,\bs\xi,Q)|\to 0 \mbox{ a.s.}
\end{equation}
The existence of a $Q$-integrable envelope of $U$ together with
continuity of $U$ imply the continuity of $V$
as function of $(\bs\lambda,\bs\theta,\bs\xi)\in{\cal C}$. This and A2 imply
$T_{n}(\tilde\Y_n)\to T(Q)$ a.s., because otherwise with probability $>0$
a subsequence of $T_{n}(\tilde\Y_n)$ can converge to $(\bs\lambda^*,\bs\theta^*,\bs\xi^*)\neq T(Q)$ but still
$\in{\cal C}$ and  with
$V(\bs\lambda^*,\bs\theta^*,\bs\xi^*,Q)=V(T(Q),Q)$, with contradiction
to A2.

\section{Experiments}\label{sexp}
\subsection{Simulation study}

We evaluated the performance of the VWQC in a simulation study under four scenarios with two populations:
\begin{enumerate}
\item observations generated independently from a multivariate distribution with Student's $t_3$ marginals;
\item observations generated as in the previous scenario, then each variable $W_j (j=1,\dots, p)$ was subsequently transformed according to $W_j \mapsto \log(|W_j|)$, to induce asymmetry;
\item observations generated from a multivariate distribution with exponential marginals;
\item observations generated  from a multivariate Gaussian distribution, but the $p$ variables $W_j$ were split into five balanced blocks to which we applied different transformations: (i) $W_j \mapsto W_j$; (ii) $W_j \mapsto \exp(W_j)$; (iii) $W_j\mapsto \log (|W_j|)$; (iv) $W_j \mapsto W_j^2$; (v) $W_j \mapsto |W_j|^{0.5}$.
\end{enumerate}

\noindent The two populations differed by a location shift equal to $+0.5$ for $Y_j$ with respect to $X_j$, $j=1,\dots,p$, for the first scenario,
$\pm 0.4$ for the second scenario (+0.4 for half of the variables, -0.4 for the other half) and $+0.2$ for the third and the fourth scenarios. For
each scenario, we evaluated combinations of overall sample sizes $n\in\{50, 100, 500\}$ (with $n/2$ observations in each class), dimensions
$p\in\{10, 50, 100, 500\}$ and percentages of relevant variables for classification in $\{10\%, 50\% , 100\%\}$. Furthermore, we considered both
uncorrelated and correlated variables. In the second case, the variance-covariance matrix was defined as in
\citep{farcomeni2022directional} by using $\mathbf{\Sigma} = \mathbf{A}^T \mathrm{diag}(\sigma_1, \sigma_2, \dots, \sigma_p)\mathbf{A}$, with
$\sigma_j = \sigma_j^*/\max(\sigma_1^*, \sigma_2^*, \dots, \sigma_p^*); \, \sigma_j^* = (p + 2 - j )^{1.1 + \frac{0.8}{p-1}j}$, and $\mathbf{A}$
being an orthogonal matrix obtained from the QR decomposition of a $p \times p$ matrix $\mathbf{B}$ defined so that $\mathrm{vec}(\mathbf{B}) = (1,
1+\frac{1}{p^2-1},\dots,1+\frac{p^2-2}{p^2-1}, 2)$. The corresponding pairwise correlations are on the interval  $(-0.25, 0.39)$ when $p = 10$,
$(-0.42, 0.62)$ when $p = 50$,  $(-0.48, 0.67)$ when $p = 100$, $(-0.63, 0.80)$ when $p = 500$. This resulted in a total of 288 different simulation
settings. Data generation under each setting was replicated 100 times. Observations in the training and test sets were generated in the same way.
Variables were standardised to unit within-class pooled variance in scenario 4, because their scales seem incompatible, whereas in datasets like
those from the other scenarios the reasons against standardization given in \citet[Section 4.1]{HV2016} may apply (imagining that such data arise in
the real world).

We compared the proposed quantile classifier with that of ten other classifiers in terms of the misclassification rate on the test data: the
directional quantile classifier (DQC) \citep{farcomeni2022directional}, the centroid classifier (Centroid) \citep{THNC02}, the median classifier
(Median) \citep{HTX09}, the original quantile classifier (OQC) \citep{HV2016}, the ensemble quantile classifier (EQC) \citep{LM20}, Fisher’s linear
discriminant analysis (LDA), the k-nearest neighbour classifier (KNN) \citep{CH67}, penalised logistic regression (PLR) \citep{PH08}, the support
vector machines (SVM) \citep{CV95,WZZ08}, the naive Bayes classifier (Bayes) \cite{HY01} and random forest (RF) \citep{ho1995random}. Tuning
parameters for PLR, KNN, and SVM where selected using cross-validation. The Galton correction was used on the OQC to reduce skewness, and the optimal
quantile was selected by minimising the error rate on the training set. All analyses were carried out in \texttt{R} version 4.1.1
\citep{RCT2021} using the following packages: \texttt{Qtools} for the directional quantile classifier; \texttt{quantileDA} for the centroid, median
and original quantile classifiers, \texttt{eqc} for the ensemble quantile classifier, \texttt{MASS} for linear discriminant analysis, the package
\texttt{class} for k-nearest neighbour classifier, \texttt{stepPlr} for penalised logistic regression, \texttt{e1071} for support vector machines and
the naive Bayes classifier, and \texttt{randomForest} for the random forest classifier. Except where indicated above, default tuning of the functions was used.

\begin{figure}[h!]
\includegraphics[scale=.85]{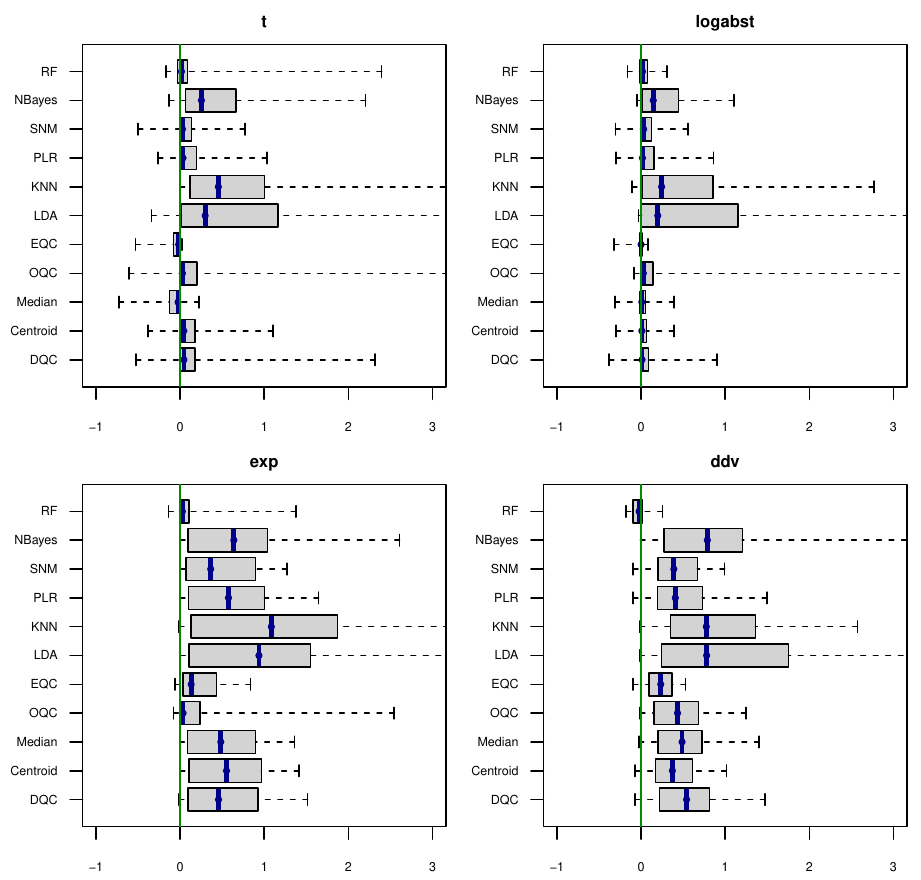}
\caption{Performance of the classifiers relative to the variable-wise quantile classifier.
The four panels show the distributions of the misclassification rates over simulated data sets for variables generated from: multivariate distributions with Student's $t_3$ marginals (t), multivariate asymmetric distributions through logarithmic (logabst), exponential distributions (exp), and different distributions (ddv).}
\label{fig:box}
\end{figure}

We computed the relative performance of each classifier with respect to the misclassification rates of the VWQC. More
specifically, we evaluated the misclassification rate of each classifier as the error rate minus the VWQC error rate divided by the average error rate in the given setting. The distributions over the 100 simulated data sets of these rescaled results for different choices of $n$,  $p$, percentage of relevant variables and dependence/independence is represented in the boxplots of Figure \ref{fig:box}.

The misclassification rates averaged over 100 replications for all simulations are reported in the Appendix in Tables \ref{t:sim t_u} to \ref{t:sim ddv_c}.
The results indicate that the performance of the VWQC improves as the sample size and the number of (relevant) variables increase: the average misclassification rate decreases and the corresponding standard errors go to zero, in agreement with the theoretical results. The VWQC outperforms its competitors with just a few exceptions. In particular, in the symmetric scenario (t), the ensemble quantile classifier and - not surprisingly - the median classifier slightly tend to prevail, as well as the random forest in the scenario with differently distributed variables (ddv). In the latter case, however, the gap between our classifier and all the other competitors is substantial.
The potential of our method is highlighted the most by the exponential scenario (exp), where even the first quartiles of the misclassification rates of any classifier relative to the VWQC are positive.

\subsection{Application to real data sets}
We applied VWQC to a collection of seven real data sets representative of a broad range of scenarios in terms of number of instances, attributes, and
classes, as described in Table \ref{t:data}. With respect to the simulation study, penalized logistic regression was not included among
the competitors due to the presence of cases with $K > 2$; the nearest shrunken centroid (NSC) \citep{THNC02} was considered instead, and applied
using the \texttt{R} package \texttt{pamr}.

\begin{table}
\caption{Short description of the data sets used for the analysis.}
\label{t:data}
\centering
\tiny
 \begin{tabular}{l  r r r c c}
 \toprule
Name		&n 	&p 	&k 	&source	&reference \\
\midrule
Bankrupt 	&66	&2	&2	&R (MixGHD)	&\citet{altman1968financial}\\
Chiro &202 &17 &5 & available upon request &\citet{atchley1971morphometric} \\
Iris & 150 & 4 & 3 &\href{https://archive.ics.uci.edu/ml/datasets/iris}{ UCI machine learning repository}, R datasets &\citet{fisher1936use}\\
Music  &306 & 20 &5 &\href{https://www.musical-style-recognition.net/?Datasets___PILOT-1.0}{musical-style-recognition.net} &\citet{van2005musical}\\
Sonar & 208 &60 & 2 &\href{http://archive.ics.uci.edu/ml/datasets/connectionist+bench+(sonar,+mines+vs.+rocks)}{ UCI machine learning repository}, R (MixGHD) &\citet{gorman1988analysis} \\
Thyroid &215 &5 &3 & \href{https://archive.ics.uci.edu/ml/datasets/thyroid+disease}{ UCI machine learning repository} & \citet{quinlan1986induction}\\
Wine &178 &13 &3 & \href{http://archive.ics.uci.edu/ml/datasets/Wine}{ UCI machine learning repository}, R(HDclassif) & \citet{aeberhard1993improvements}\\
\bottomrule
\end{tabular}
\end{table}

According to the resulting misclassification rates reported in Table \ref{t:rda}, the VWQC provided good performances, ranking always above the median and thus highlighting again the adaptability of the classifier. Furthermore, an optimal outcome is obtained for the ``bankrupt'' data set, tied with the random forest classifier. On the contrary, the highest misclassification rates for this data set are provided by the centroid classifier, NSC, and LDA. Although the latter ranks below the median in another case (``thyroid"), it shows also some highs, ranking first in three data sets (``iris", ``music", ``wine"), hence being ranked among the most successful competitors for this study, together with the random forest.

\begin{table}
\caption{Ten-fold cross-validated misclassification rates (\%) for the seven data sets, with standard errors in parentheses. For each data set, the best performances are highlighted in bold. In case of ties, those associated to the lowest standard error are selected.}
\label{t:rda}
\centering
\scriptsize
 \begin{tabular}{l  r r r r r r r}
 \toprule
    		&Bankrupt	 	&Chiro 		&Iris  			&Music			&Sonar 			&Thyroid		&Wine \\
\midrule
VWQC		&\textbf{3.1 (6.5)}	&2.5 (3.5)		&4.0 (5.6)		&32.9 (8.6)			&22.4 (12.1)			&6.0 (5.3)		&2.3 (3.0)\\
LDA   	 	&8.1 (8.6)		&1.5 (2.3)		&\textbf{2.0 (4.5)}	&\textbf{24.1 (6.7)}	&25.4 (8.0)			&8.9 (3.5)		&\textbf{1.1 (2.4)}\\
NSC    	&8.1 (8.6)		&6.5 (4.1)		&6.7  (4.4)		&38.5 (5.0)			&25.4 (9.1)			&16.7 (4.5)		&4.5 (3.7)\\
KNN       	&4.8 (7.7)		&6.9 (2.5)		&2.7  (3.4)		&39.6 (6.0)			&20.1 (7.4)			&5.2 (4.1)		&23.5 (10.2)\\
SVM       	&4.3 (7.1)		&1.5 (2.4)		&4.0  (6.4)		&24.1 (7.4)			&15.3 (5.6)			&4.7 (3.9)		&2.2 (3.8)\\
%CART     	&7.9 (8.3)		&6.9 (4.7)		&6.0  (3.8)		&36.5 (8.0)			&24.9 (11.3)			&7.8 (6.4)		&11.8 (5.5)\\
nBayes     	&6.2 (8.0)		&\textbf{1.0 (2.1)}	&4.0  (4.7)		&29.7 (4.7)			&22.9 (11.4)			&\textbf{2.8 (3.3)}	&2.8 (3.8)\\
QC       	&6.4 (8.3)		&8.9 (6.0)		&7.3  (4.9)		&43.4 (6.0)			&30.2 (8.3)			&9.3 (4.5)		&18.0 (6.3)\\
Centroid 	&9.3 (8.2)		&12.9 (6.3)		&7.3  (6.6)		&47.6 (7.3)			&31.7 (5.7)			&13.5 (7.4)		&26.4 (9.3)\\
Median   	&3.3 (7.0)		&9.0 (4.6)		&8.0  (5.3)		&42.1 (5.5)			&28.3 (9.9)			&11.2 (7.3)		&26.9 (9.1)\\
EQC    	&6.2 (8.0)		&4.0 (4.6)		&5.3  (4.2)		&34.0 (7.0)			&22.5 (9.2)			&4.6 (4.3)		&3.4 (2.9)\\
RF       	&\textbf{3.1 (6.5)}	&\textbf{1.0 (2.1)}	&4.7  (5.5)		&29.0 (7.2)			&\textbf{13.4 (9.7)}	&3.7 (2.9)		&1.7 (2.7)\\
DQC		&3.3 (7.0)		&12.9 (4.2)		&7.3  (4.9)		&55.8 (3.7)			&33.0 (9.2)			&11.2 (6.7)		&26.3 (8.3)\\
\bottomrule
\end{tabular}
\end{table}

\section{Conclusion}\label{sconc}
It can be expected that in most high-dimensional classification problems different variables have different distributional shapes within classes. The original quantile classifier by \cite{HV2016} is based on just a single percentage value for computing the quantiles used for classification of all variables. Although this often works well and a single parameter is easy to optimise, it can be expected that using different percentage parameters for different variables may use the information in the data more efficiently. This requires a rescaling of the quantile-based discrepancy, and because of this, we introduce variable-wise scaling parameters on top of the percentages. The resulting variable-wise quantile classifier is still relatively simple to compute, and in many situations it can improve upon the original quantile classifier. Its performance, while not universally optimal, also can compete well with other existing popular classifiers.

Although the VWQC can be extended to more than two classes in a straightforward manner, we constrained ourselves to two classes in the simulation study. Generally, there are lots of options and possibilities for such studies so that certain limitations of this kind are inevitable. Most of the considered real data sets  come with more than two classes.

Interesting further work could be done by exploring possibilities for a theory for high-dimensional settings, i.e., allowing $p\to\infty$, and for finite sample performance guarantees.

\bibliographystyle{chicago}
\bibliography{quantiles_0623_arXiv}

\appendix
\section{Simulation study -- misclassification rates}

% latex table generated in R 3.6.3 by xtable 1.8-4 package
% Tue Aug 24 19:02:48 2021
\begin{sidewaystable}[ht]
\centering
\caption{Misclassification rates (with standard deviations) averaged  over 100 replications in brackets for twelve classifiers in the t scenario with uncorrelated variables.}
\label{t:sim t_u}
\begingroup \tiny%\scriptsize
\begin{tabular}{l ccc|ccc|ccc}
  \toprule
Sample size $n$ & 50 & 100 & \multicolumn{1}{c}{500}  & 50 & 100 & \multicolumn{1}{c}{500} & 50 & 100 & 500  \\
  \midrule
Relevant variables & & 10\% & & & 50\% & & & 100\% & \\
  \midrule
& \multicolumn{9}{c}{$p=10$}\\
\midrule
VWQC & 0.436 (0.048) & 0.455 (0.033) & 0.426 (0.025) & 0.436 (0.046) & 0.442 (0.038) & 0.433 (0.028) & 0.443 (0.044) & 0.450 (0.036) & 0.426 (0.028) \\
  DQC & 0.443 (0.046) & 0.458 (0.031) & 0.467 (0.020) & 0.440 (0.045) & 0.457 (0.032) & 0.471 (0.022) & 0.443 (0.044) & 0.455 (0.034) & 0.462 (0.021) \\
  Centroid & 0.439 (0.049) & 0.453 (0.037) & 0.445 (0.023) & 0.447 (0.037) & 0.455 (0.038) & 0.445 (0.023) & 0.439 (0.042) & 0.453 (0.036) & 0.443 (0.025) \\
  Median & 0.437 (0.047) & 0.453 (0.034) & 0.422 (0.025) & 0.440 (0.045) & 0.441 (0.042) & 0.429 (0.025) & 0.438 (0.045) & 0.439 (0.043) & 0.422 (0.025) \\
  CQC & 0.445 (0.045) & 0.457 (0.036) & 0.435 (0.028) & 0.448 (0.044) & 0.449 (0.037) & 0.438 (0.026) & 0.442 (0.047) & 0.454 (0.037) & 0.435 (0.028) \\
\bottomrule
\end{tabular}
\endgroup
%\end{adjustbox}
\end{sidewaystable}

% latex table generated in R 4.2.1 by xtable 1.8-4 package
% Tue Dec 27 23:55:33 2022
\begin{sidewaystable}[ht]
\centering
\caption{Misclassification rates (with standard deviations) averaged  over 100 replications in brackets for twelve classifiers in the t scenario with correlated variables.}
\label{t:sim t_c}
\begingroup\tiny
\begin{tabular}{l ccc|ccc|ccc}
  \toprule
Sample size $n$ & 50 & 100 & \multicolumn{1}{c}{500}  & 50 & 100 & \multicolumn{1}{c}{500} & 50 & 100 & 500  \\
  \midrule
Relevant variables & & 10\% & & & 50\% & & & 100\% & \\
  \midrule
& \multicolumn{9}{c}{$p=10$}\\
\midrule
VWQC & 0.429 (0.049) & 0.438 (0.042) & 0.414 (0.031) & 0.439 (0.043) & 0.436 (0.042) & 0.411 (0.027) & 0.434 (0.050) & 0.443 (0.042) & 0.412 (0.026) \\
  DQC & 0.441 (0.047) & 0.451 (0.035) & 0.455 (0.024) & 0.442 (0.050) & 0.456 (0.035) & 0.457 (0.020) & 0.437 (0.048) & 0.450 (0.034) & 0.454 (0.025) \\
  Centroid & 0.441 (0.046) & 0.443 (0.037) & 0.424 (0.026) & 0.443 (0.043) & 0.449 (0.042) & 0.424 (0.023) & 0.438 (0.053) & 0.439 (0.042) & 0.427 (0.026) \\
  Median & 0.435 (0.051) & 0.438 (0.041) & 0.408 (0.028) & 0.437 (0.050) & 0.438 (0.047) & 0.411 (0.027) & 0.432 (0.049) & 0.438 (0.046) & 0.411 (0.027) \\
  CQC & 0.438 (0.049) & 0.447 (0.039) & 0.420 (0.029) & 0.445 (0.039) & 0.442 (0.040) & 0.423 (0.029) & 0.435 (0.049) & 0.445 (0.044) & 0.419 (0.027) \\
\bottomrule
\end{tabular}
\endgroup
\end{sidewaystable}

% latex table generated in R 4.2.1 by xtable 1.8-4 package
% Tue Dec 27 23:55:33 2022
\begin{sidewaystable}[ht]
\centering
\caption{Misclassification rates (with standard deviations) averaged  over 100 replications in brackets for twelve classifiers in the logabst scenario with uncorrelated variables.}
\label{t:sim logabst_u}
\begingroup\tiny
\begin{tabular}{l ccc|ccc|ccc}
  \toprule
Sample size $n$ & 50 & 100 & \multicolumn{1}{c}{500}  & 50 & 100 & \multicolumn{1}{c}{500} & 50 & 100 & 500  \\
  \midrule
Relevant variables & & 10\% & & & 50\% & & & 100\% & \\
  \midrule
& \multicolumn{9}{c}{$p=10$}\\
\midrule
VWQC & 0.442 (0.047) & 0.459 (0.035) & 0.441 (0.027) & 0.446 (0.044) & 0.449 (0.037) & 0.439 (0.023) & 0.441 (0.043) & 0.444 (0.041) & 0.438 (0.025) \\
  DQC & 0.439 (0.049) & 0.456 (0.031) & 0.467 (0.019) & 0.451 (0.043) & 0.450 (0.035) & 0.464 (0.019) & 0.447 (0.043) & 0.454 (0.029) & 0.462 (0.020) \\
  Centroid & 0.439 (0.042) & 0.450 (0.040) & 0.447 (0.024) & 0.449 (0.047) & 0.449 (0.037) & 0.444 (0.021) & 0.439 (0.049) & 0.455 (0.034) & 0.444 (0.024) \\
  Median & 0.439 (0.048) & 0.453 (0.037) & 0.443 (0.027) & 0.443 (0.040) & 0.448 (0.042) & 0.441 (0.022) & 0.449 (0.036) & 0.457 (0.036) & 0.443 (0.026) \\
  CQC & 0.445 (0.046) & 0.455 (0.036) & 0.446 (0.029) & 0.441 (0.046) & 0.447 (0.039) & 0.446 (0.026) & 0.445 (0.046) & 0.453 (0.037) & 0.442 (0.027) \\
\bottomrule
\end{tabular}
\endgroup
\end{sidewaystable}

% latex table generated in R 4.2.1 by xtable 1.8-4 package
% Tue Dec 27 23:55:33 2022
\begin{sidewaystable}[ht]
\centering
\caption{Misclassification rates (with standard deviations) averaged  over 100 replications in brackets for twelve classifiers in the logabst scenario with correlated variables.}
\label{t:sim logabst_c}
\begingroup\tiny
\begin{tabular}{l ccc|ccc|ccc}
  \toprule
Sample size $n$ & 50 & 100 & \multicolumn{1}{c}{500}  & 50 & 100 & \multicolumn{1}{c}{500} & 50 & 100 & 500  \\
  \midrule
Relevant variables & & 10\% & & & 50\% & & & 100\% & \\
  \midrule
& \multicolumn{9}{c}{$p=10$}\\
\midrule
VWQC & 0.431 (0.056) & 0.445 (0.037) & 0.440 (0.025) & 0.439 (0.045) & 0.450 (0.040) & 0.434 (0.026) & 0.435 (0.055) & 0.447 (0.038) & 0.440 (0.023) \\
  DQC & 0.449 (0.041) & 0.450 (0.034) & 0.469 (0.018) & 0.436 (0.046) & 0.459 (0.032) & 0.469 (0.020) & 0.436 (0.051) & 0.450 (0.039) & 0.467 (0.021) \\
  Centroid & 0.445 (0.045) & 0.445 (0.038) & 0.448 (0.023) & 0.437 (0.047) & 0.453 (0.033) & 0.447 (0.024) & 0.433 (0.050) & 0.449 (0.036) & 0.449 (0.024) \\
  Median & 0.438 (0.050) & 0.449 (0.037) & 0.439 (0.026) & 0.448 (0.042) & 0.448 (0.036) & 0.442 (0.026) & 0.439 (0.045) & 0.453 (0.033) & 0.443 (0.025) \\
  CQC & 0.440 (0.046) & 0.453 (0.040) & 0.444 (0.026) & 0.443 (0.039) & 0.443 (0.039) & 0.445 (0.025) & 0.432 (0.049) & 0.455 (0.034) & 0.445 (0.027) \\
\bottomrule
\end{tabular}
\endgroup
\end{sidewaystable}

% latex table generated in R 4.2.1 by xtable 1.8-4 package
% Tue Dec 27 23:55:34 2022
\begin{sidewaystable}[ht]
\centering
\caption{Misclassification rates (with standard deviations) averaged  over 100 replications in brackets for twelve classifiers in the exp scenario with uncorrelated variables.}
\label{t:sim exp_u}
\begingroup\tiny
\begin{tabular}{l ccc|ccc|ccc}
  \toprule
Sample size $n$ & 50 & 100 & \multicolumn{1}{c}{500}  & 50 & 100 & \multicolumn{1}{c}{500} & 50 & 100 & 500  \\
  \midrule
Relevant variables & & 10\% & & & 50\% & & & 100\% & \\
  \midrule
& \multicolumn{9}{c}{$p=10$}\\
\midrule
VWQC & 0.432 (0.051) & 0.435 (0.046) & 0.420 (0.020) & 0.444 (0.043) & 0.436 (0.045) & 0.420 (0.021) & 0.441 (0.045) & 0.436 (0.041) & 0.422 (0.017) \\
  DQC & 0.450 (0.036) & 0.456 (0.030) & 0.476 (0.017) & 0.439 (0.046) & 0.457 (0.031) & 0.473 (0.018) & 0.444 (0.039) & 0.466 (0.026) & 0.476 (0.017) \\
  Centroid & 0.444 (0.042) & 0.460 (0.030) & 0.468 (0.019) & 0.446 (0.041) & 0.459 (0.030) & 0.468 (0.021) & 0.441 (0.048) & 0.462 (0.029) & 0.472 (0.021) \\
  Median & 0.449 (0.045) & 0.461 (0.030) & 0.467 (0.022) & 0.441 (0.043) & 0.458 (0.029) & 0.466 (0.023) & 0.441 (0.047) & 0.459 (0.030) & 0.470 (0.019) \\
  CQC & 0.440 (0.043) & 0.448 (0.036) & 0.435 (0.027) & 0.446 (0.040) & 0.445 (0.041) & 0.433 (0.025) & 0.437 (0.045) & 0.454 (0.035) & 0.437 (0.025) \\
\bottomrule
\end{tabular}
\endgroup
\end{sidewaystable}

% latex table generated in R 4.2.1 by xtable 1.8-4 package
% Tue Dec 27 23:55:34 2022
\begin{sidewaystable}[ht]
\centering
\caption{Misclassification rates (with standard deviations) averaged  over 100 replications in brackets for twelve classifiers in the exp scenario with correlated variables.}
\label{t:sim exp_c}
\begingroup\tiny
\begin{tabular}{l ccc|ccc|ccc}
  \toprule
Sample size $n$ & 50 & 100 & \multicolumn{1}{c}{500}  & 50 & 100 & \multicolumn{1}{c}{500} & 50 & 100 & 500  \\
  \midrule
Relevant variables & & 10\% & & & 50\% & & & 100\% & \\
  \midrule
& \multicolumn{9}{c}{$p=10$}\\
\midrule
VWQC & 0.433 (0.052) & 0.442 (0.036) & 0.431 (0.018) & 0.431 (0.048) & 0.442 (0.038) & 0.434 (0.022) & 0.440 (0.052) & 0.445 (0.037) & 0.432 (0.021) \\
  DQC & 0.445 (0.045) & 0.456 (0.033) & 0.476 (0.017) & 0.436 (0.045) & 0.457 (0.034) & 0.472 (0.019) & 0.450 (0.037) & 0.458 (0.033) & 0.471 (0.019) \\
  Centroid & 0.442 (0.044) & 0.459 (0.034) & 0.468 (0.020) & 0.443 (0.041) & 0.453 (0.033) & 0.462 (0.020) & 0.441 (0.045) & 0.454 (0.034) & 0.462 (0.021) \\
  Median & 0.436 (0.045) & 0.460 (0.034) & 0.468 (0.020) & 0.444 (0.045) & 0.458 (0.035) & 0.464 (0.023) & 0.446 (0.046) & 0.454 (0.032) & 0.463 (0.023) \\
  CQC & 0.441 (0.047) & 0.451 (0.038) & 0.438 (0.023) & 0.438 (0.051) & 0.453 (0.036) & 0.440 (0.022) & 0.442 (0.047) & 0.450 (0.040) & 0.441 (0.024) \\
\bottomrule
\end{tabular}
\endgroup
\end{sidewaystable}

% latex table generated in R 4.2.1 by xtable 1.8-4 package
% Tue Dec 27 23:55:34 2022
\begin{sidewaystable}[ht]
\centering
\caption{Misclassification rates (with standard deviations) averaged  over 100 replications in brackets for twelve classifiers in the ddv scenario with uncorrelated variables.}
\label{t:sim ddv_u}
\begingroup\tiny
\begin{tabular}{l ccc|ccc|ccc}
  \toprule
Sample size $n$ & 50 & 100 & \multicolumn{1}{c}{500}  & 50 & 100 & \multicolumn{1}{c}{500} & 50 & 100 & 500  \\
  \midrule
Relevant variables & & 10\% & & & 50\% & & & 100\% & \\
  \midrule
& \multicolumn{9}{c}{$p=10$}\\
\midrule
VWQC & 0.431 (0.045) & 0.459 (0.032) & 0.468 (0.021) & 0.357 (0.067) & 0.318 (0.050) & 0.301 (0.022) & 0.261 (0.067) & 0.228 (0.048) & 0.204 (0.021) \\
  DQC & 0.444 (0.043) & 0.455 (0.034) & 0.472 (0.019) & 0.415 (0.055) & 0.419 (0.047) & 0.404 (0.023) & 0.374 (0.066) & 0.359 (0.049) & 0.339 (0.019) \\
  Centroid & 0.443 (0.047) & 0.452 (0.037) & 0.469 (0.020) & 0.411 (0.062) & 0.404 (0.049) & 0.384 (0.025) & 0.370 (0.072) & 0.352 (0.047) & 0.330 (0.022) \\
  Median & 0.445 (0.046) & 0.451 (0.038) & 0.471 (0.022) & 0.415 (0.054) & 0.405 (0.050) & 0.392 (0.023) & 0.380 (0.064) & 0.373 (0.049) & 0.351 (0.022) \\
  CQC & 0.445 (0.042) & 0.455 (0.038) & 0.474 (0.019) & 0.408 (0.061) & 0.381 (0.059) & 0.345 (0.046) & 0.341 (0.078) & 0.287 (0.077) & 0.258 (0.069) \\
\bottomrule
\end{tabular}
\endgroup
\end{sidewaystable}

% latex table generated in R 4.2.1 by xtable 1.8-4 package
% Tue Dec 27 23:55:34 2022
\begin{sidewaystable}[ht]
\centering
\caption{Misclassification rates (with standard deviations) averaged  over 100 replications in brackets for twelve classifiers in the t scenario with correlated variables.}
\label{t:sim ddv_c}
\begingroup\tiny
\begin{tabular}{l ccc|ccc|ccc}
  \toprule
Sample size $n$ & 50 & 100 & \multicolumn{1}{c}{500}  & 50 & 100 & \multicolumn{1}{c}{500} & 50 & 100 & 500  \\
  \midrule
Relevant variables & & 10\% & & & 50\% & & & 100\% & \\
  \midrule
& \multicolumn{9}{c}{$p=10$}\\
\midrule
VWQC & 0.433 (0.048) & 0.458 (0.033) & 0.469 (0.019) & 0.333 (0.066) & 0.305 (0.053) & 0.278 (0.017) & 0.222 (0.058) & 0.193 (0.043) & 0.175 (0.020) \\
  DQC & 0.443 (0.040) & 0.456 (0.033) & 0.474 (0.019) & 0.410 (0.060) & 0.415 (0.046) & 0.387 (0.022) & 0.351 (0.068) & 0.336 (0.049) & 0.323 (0.022) \\
  Centroid & 0.441 (0.042) & 0.453 (0.036) & 0.464 (0.021) & 0.396 (0.059) & 0.394 (0.052) & 0.363 (0.020) & 0.344 (0.068) & 0.328 (0.048) & 0.311 (0.024) \\
  Median & 0.439 (0.048) & 0.454 (0.035) & 0.470 (0.020) & 0.401 (0.064) & 0.405 (0.050) & 0.370 (0.021) & 0.362 (0.069) & 0.348 (0.050) & 0.329 (0.022) \\
  CQC & 0.438 (0.042) & 0.455 (0.037) & 0.472 (0.019) & 0.387 (0.072) & 0.354 (0.069) & 0.320 (0.044) & 0.289 (0.081) & 0.241 (0.071) & 0.209 (0.061) \\
\bottomrule
\end{tabular}
\endgroup
\end{sidewaystable}

\end{document}